**Lithium plating induced degradation during fast charging of batteries subjected to compressive loading.**

Prashant P. Gargh, Abhishek Sarkar, Ikenna C. Nlebedim and Pranav Shrotriya

Department of Mechanical Engineering, Iowa State University

Ames National Laboratory of US Department of Energy

Ames, Iowa 50011 USA

# ABSTRACT

We report the lithium plating associated capacity loss during fast charging of compressively loaded lithium-ion batteries (LIBs). The charging and discharging of LIB under compressive loading during service may affect the cell performance or initiate localized defects in the electrodes. Pouch cells of capacity 20mAh were compressively loaded to nominal pressures of 0-440 kPa and subjected to 10 cycles of fast charging at 1C and 4C. Experimental results show that cells charged at 4C-rate experienced significant capacity fade, and applying compressive loads exacerbated the capacity loss. The coulombic efficiency study shows that active lithium loss was higher for the initial cycles before gradually reducing to a minimal capacity loss for the tenth charging cycle. The cell voltage relaxation immediately after charging was monitored to identify the stripping of plated lithium after fast charging cycles and showed that the duration of lithium stripping was higher for cells under mechanical compressive loading. Scanning electron microscopy (SEM) and Electron paramagnetic resonance spectroscopy (EPR) characterization of the anode showed significantly higher lithium deposits on the anodes charged at a 4C rate under compressive loads. These results indicate that applied mechanical compression causes increased lithium plating during fast charging of batteries.

**HIGHLIGHTS:**

- Cells charge-cycled at high C rates under mechanical pressure show an increase in capacity loss
- Cells charged-cycled at a high C-rate under mechanical pressure show increased stripping duration immediately after charging.
- EPR and SEM imaging show enhanced lithium deposition on anodes from compressive-loaded cells.
- Application of mechanical pressure increases lithium plating in lithium-ion pouch cells.

# INTRODUCTION:

Lithium-ion batteries (LIBs) are used in portable electronics and electric vehicles (EVs) for energy storage due to their high energy and power densities, high cycle life, and low cost [1, 2]. The LIBs may also satisfy the increased need for stationary energy storage for power utilities because of their high energy density. However, the battery environment and charging protocol must be managed for safe and reliable performance. Fast charging of LIBs is required to reduce charging times in electric vehicles but can impact the battery service life, increase degradation rate, and adversely impact safety [3, 4]. Batteries operating under compressive loads may lead to high capacity loss, power fade, safety and integrity concerns [5-7].

Battery operation under–compressive loading may reduce performance and mechanical integrity concerns. Safe and reliable operation of LIBs requires understanding the influence of mechanical loads generated from extrinsic and intrinsic factors [8, 9]. A recent review has summarized the influence of extrinsic and intrinsic mechanical loads on battery performance and reliability [10]. We provide a summary of the relevant studies on how battery performance is influenced by extrinsic factors such as rigid housing constraints [11, 12], vibrations & shocks [13], stack pressure [9], as well as intrinsic factors such as charging-induced expansion [14], gas evolution and, temperature changes [15]. Previous studies have shown that applying low magnitude but optimal mechanical pressure can help increase their cycling performance, leading to an increased lifetime of the LIBs [11, 12, 16, 17]. Bach et al. [18] suggested utilizing homogeneous pressure distribution on the lithium-ion battery can improve cycle life. Muller et al. [19] demonstrated that applying low mechanical pressure improves the electrical contact in LIBs because of compression experienced by the separator and anode. Applying moderate mechanical pressure may result in faster kinetics and better performance for lithium cobalt oxide (LCO) electrodes because mechanical pressure causes a reduction in the pore volume [20].

However, Gnanaraj et al. (2001) showed that applying compressive load above the mechanical strength of electrode particles may result in the degradation of conventional graphite/LCO batteries. The LCO particles are more rigid than graphite flakes, and applying compressive load can cause localized densification, reducing lithium insertion sites and affecting their cyclability performance. Peabody et al. (2011) demonstrated that with increased compressive stress, the separator experiences viscoelastic creep, causing a reduction of lithium-ion transport due to pore closure, resulting in capacity fade. Sarkar et al. (2018) have modeled the influence of uniform compressive loading on the separator and showed that the reduced ionic conductivity due to decreased permeability might result in capacity loss and runaway heating under high pressures.

In rechargeable lithium metal batteries, moderate stack pressure may address the two main challenges of low coulombic efficiency and spatially nonuniform lithium deposition on the anode surface during charging at high rates (C-rates) [16, 21]. Fang et al. [22] reported that applying uniaxial stack pressure with an increasing current rate on lithium copper cells caused dense lithium deposition in columnar structures, increasing coulombic efficiency. The uncontrolled lithium deposits may be suppressed and further tuned into columnar structures, which could help reduce lithium inventory loss by reducing the electrode exposure area with electrolytes. Therefore, mechanical compression during charging may limit uncontrolled lithium deposits in rechargeable lithium metal batteries [22, 23].

Graphite anode-based lithium battery packs widely used in electric vehicles are subjected to stack pressure during assembly and may be subjected to high mechanical load in case of an accident. Therefore, it is important to understand and investigate the influence of compressive loading on the fast-charging performance of batteries. We report the effect of uniform compressive load on the capacity fade during the cycling of LIBs at fast charging rates. Commercial graphite/LCO battery performance was subjected to four different pressure levels between 0 - 440 KPa at two charging rates, 1C and 4C. The cell voltage relaxation immediately after charging was monitored to identify lithium plating. After the cycling experiments, the batteries were discharged completely and safely dissembled, and post-mortem characterizations were performed to identify the influence of compressive loading on the battery electrodes. Electron paramagnetic resonance (EPR) spectroscopy measurements were carried out to detect the metallic lithium content on the graphite anode. Scanning electron microscopy (SEM) imaging and Energy Dispersive X-ray (EDX) spectroscopy were utilized to characterize the influence of fast charging under compressive loading on the changes in surface morphology and chemical composition of the anode surfaces.

## 2. MATERIALS AND METHODS:

### 2.1: Pouch cells:

All experiments were conducted on commercial lithium cobalt oxide/ graphite (LCO/G) pouch cells (PowerStream Technologies, model number GM201515) with a nominal capacity of 20 mAh, an operating voltage range of 3.0 V minimum and 4.2V maximum, and dimensions of 2 mm × 15 mm × 15 mm.

### 2.2 Compressive Mechanical Loading Fixture:

The pouch cells were loaded to four different pressure levels of 0KPa, 40 KPa, 200 KPa, and 440 KPa using the loading setup shown in Figure 1(a) and subjected to charge-discharge cycling at different C-rates for ten cycles. The pouch cells were compressed to the desired nominal pressure level, and the magnitude of the applied load was monitored at a 1 Hz sampling rate during the charge-discharge cycles.

### 2.3: Charging Protocol:

The charge/discharge cycle protocol for the cells is shown in Figure 1(b). The initial charge capacity of the pouch cells was measured using an MTI BST8-300-CST battery cycler by charging at a C/10 rate to a cell voltage of 4.2V. The cells were rested for 30 minutes, and the discharge capacity of the cell was measured by discharging the cell to 3.0 V at a C/10 rate.

After the initial capacity measurement, the cells were loaded to the desired loading level and cycled for ten cycles using Gamry Ref 600 potentiostat. The cells were charged at the desired rate (1C or 4C) in each fast-charging cycle until 90% of the measured capacity. If the cell voltage reached 4.2 V before 90% capacity, the cells were charged under constant voltage (CV) till the charging current decays to C/2. After each charging cycle, the cells were rested for 30 minutes and discharged to 3.0V at C/2 rate. The discharged cells rested again for 30 minutes before the next charging cycle.

After ten cycles of fast charging, the cells were removed from the loading fixture, and the final capacity of the cells was measured using the same protocol as the initial capacity measurement.

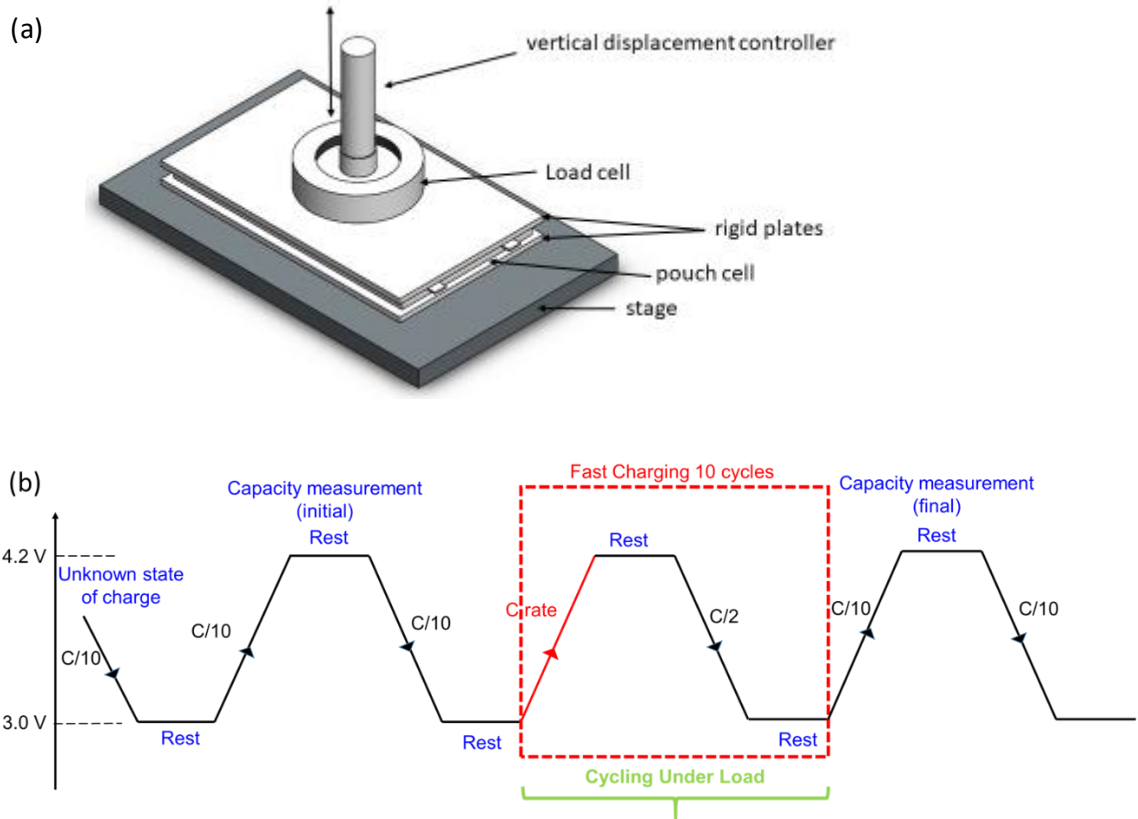

Figure 1: (a) Schematic for compressive loading setup designed for mechanical pressure application on lithium-ion pouch cells (b): Cycling protocol for chosen C rate

**2.4 Postmortem Analysis:**

All charge-cycled pouch cells were discharged to 2.7 V and transferred into an MBraun glovebox (< 0.1 ppm of $H_2O$ and $O_2$). The cell casing was removed, and the anode/separator/cathode roll was carefully unfolded to avoid damaging the film deposited on the anode/separator interface. The anode samples were transferred from the glovebox to the SEM stage using a vacuum sample holder, and SEM imaging was performed using FEI Teneo to study the surface morphology evolution. The extracted anode was cut into five 2 mm strips and sealed in an EPR tube (Suprasil, $\phi$ 6 mm) using silicone septum in the glove box and transferred to ELEXSYS E580 FT-EPR for EPR characterization to detect bulk Li metal deposition.

## 3. RESULTS & DISCUSSIONS:

### 3.1: Capacity evolution during cycling:

The measured columbic efficiency (CE) (ratio of the Li$^+$ ions extracted during discharge to the number of Li$^+$ ions inputted during charge [24]) of the cells charged at 1C and 4C charging rates are plotted as a function of charging cycles in Figures 2(a) and 2(b), respectively. The columbic efficiency under the two charging rates was measured for four different compressive loads (0 KPa (no pressure), 40 KPa, 200 KPa, and 440 KPa). Measured discharge capacity corresponding to 1C and 4C charging rates are included in the supplementary information (Figures S1(a) and S1(b), respectively).

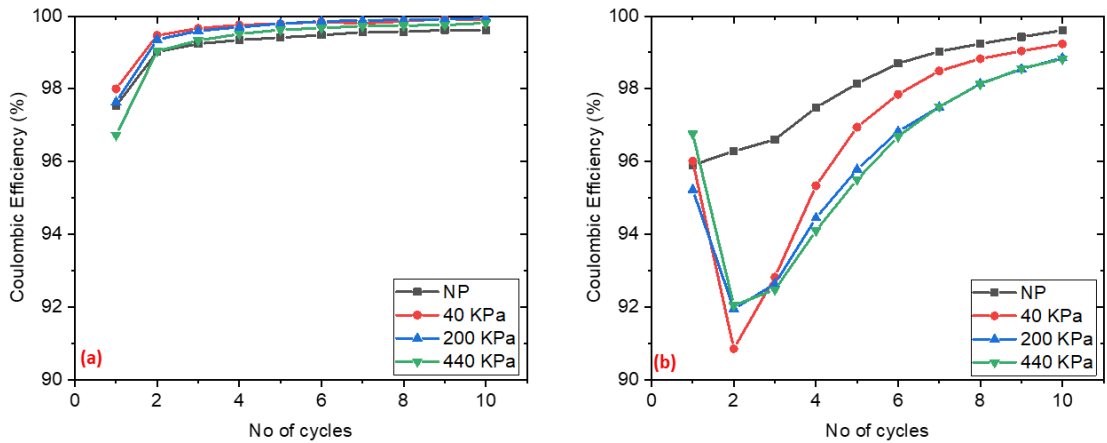

Figure 2: Coulombic efficiency vs. the number of cycles for C rate (a) 1C and (b) 4C, respectively, as a function of applied external pressure.

The CE in the cells charged at the 1C rate drops during the first charging cycle but recovers to nearly 99% during subsequent cycles, as shown in Figure 2(a), and minimal discharge capacity beyond the first cycle. The CE and discharge capacity of the cells cycled under all the compressive loads shows a similar behavior as the unloaded cells for the 1C rate. The CE values lower than 100% indicate a loss of active lithium inventory [24-26] due to parasitic reactions associated with either SEI reformation/growth on the anode surface and/or lithium plating during charging [27]. Measured CE values for cycling at 1C indicate that the first charging cycle results in lithium inventory loss, but the losses do not accumulate during subsequent cycling. Applying compressive loads does not influence the charging at 1C over the load range and charging cycles tested in these experiments.

The CE of the cells cycled at 4C rates (shown in Fig 2(b)) is lower for all ten cycles than for cells cycled at 1C. Similarly, the loss in discharge capacity in all cells at cycles at 4C (plotted in Fig S1(b)) is greater than in cells cycled at 1C. Discharge capacity (plotted in Fig S1(b)) decreases rapidly for the first few cycles, but the incremental loss per cycle decreases with subsequent cycles. The results show that loss in lithium inventory due to parasitic reactions is greater at 4C charge rates. The application of compressive loads exacerbates the charge efficiency and capacity losses. The cells subjected to compressive loads show a larger drop in CE and a decrease in discharge capacity during the cycling compared to unloaded cells.

Cell voltage evolution during the rest periods immediately after charging at 4C is compared for all ten cycles of cells subjected to no load and maximum compressive loading (440 KPa) in Figures 3 (a) and 3 (b), respectively. Cell voltages measured for other loading cases are included in the supplementary information (Fig S2 for 40 KPa and Fig S3 for 200 KPa). The open circuit cell voltage evolution during relaxation is attributed to lithium diffusion and parasitic reactions inside the cell [25, 28, 29]. During the initial cycles of the unloaded cell (Figs 3(a)), cell voltage decays rapidly but reaches an intermediate plateau region before declining to a final stable value. Additionally, the duration of the intermediate plateau region in the voltage evolution becomes shorter with an increase in cycles. For the final cycles, the voltage evolution of the cells follows a smooth exponential decay and reaches a stable value after approximately 10 min of relaxation. The voltage evolution of the loaded cells follows a similar trend during cycling as unloaded cells, but the duration of the intermediate plateau becomes longer with an increase in compressive pressure, and the intermediate plateau is observed for more cycles. For unloaded pouch cells, the voltage plateau fades away after four cycles (Fig 3(a)), whereas for pouch cells under 440 KPa, this voltage plateau fades away after about seven cycles (Fig 3(b)). The intermediate plateau region is referred to as the region of mixed potential, i.e., the potential between intercalated lithium (in graphite) and plated lithium [30, 31], and indicates stripping of lithium plated during the charging step [32].

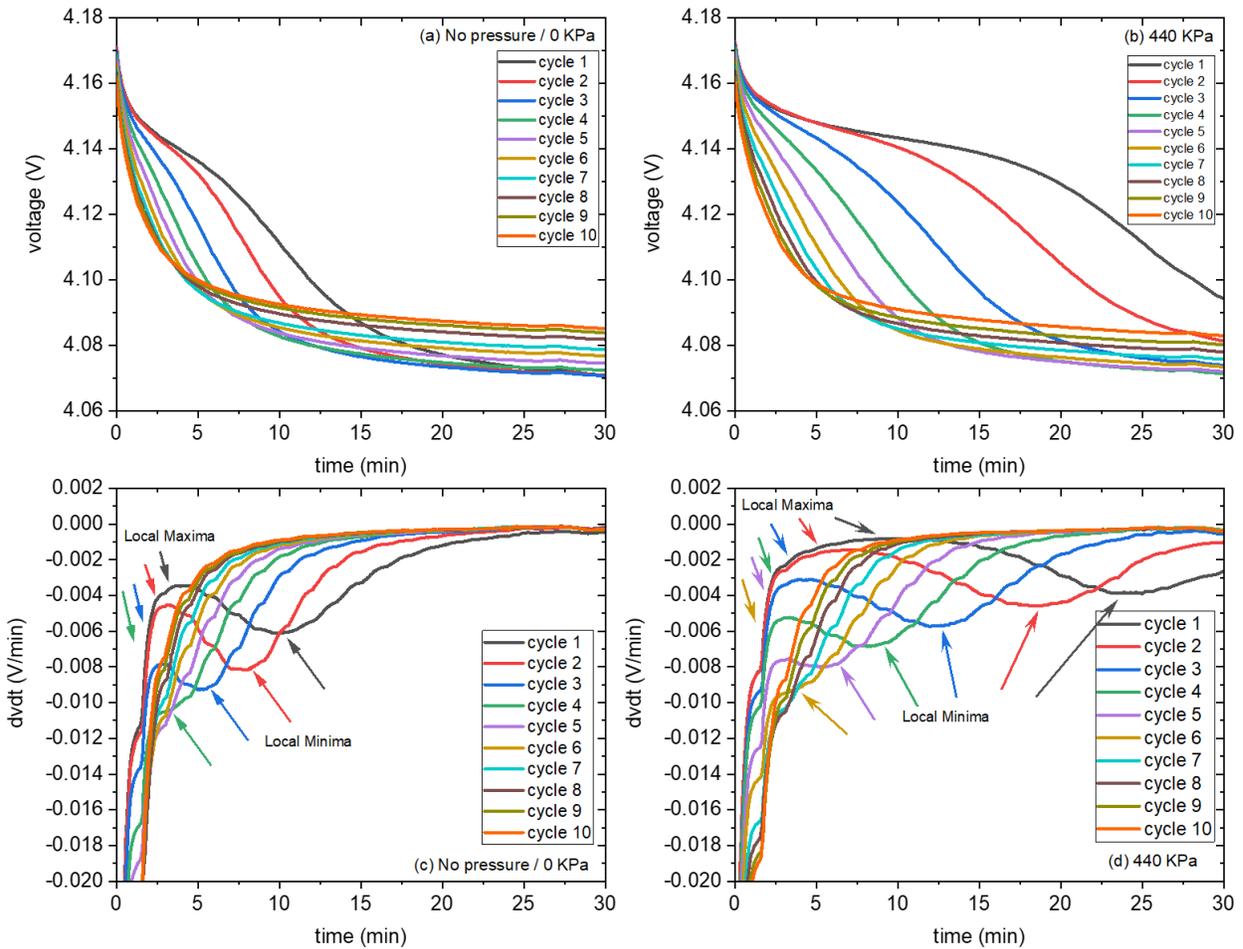

Figure 3: Voltage relaxation and differential voltage curves for cells charged with 4C charging rate under (a),(c) NP (b),(d) 0.44 MPa.

The rate of voltage decay (dv/dt) calculated from data in figures 3(a) and 3(b) are plotted in figures 3(c) and 3(d), respectively, and show local maxima and minima that correspond to the beginning and end of the plateaus observed in figures 3(a) and 3(b) during relaxation. The local maxima in the rate of voltage decay may indicate the beginning of stripping of deposited lithium, and the local minima correspond to the end of stripping [33] [31, 34].

The duration of stripping for the cells cycled at 4C was estimated from the rate of voltage decay curves and is plotted in Figure 4(a) as a function of cycles for the cells loaded to the four loading levels. The stripping duration was maximum for the first cycle and decreased with cycles. Applying compressive loads increased the stripping duration, as shown in Figure 4(a). During the first cycle, the stripping duration under 440KPa pressure was almost 2.5 times longer than the 0KPa (no pressure) condition (Figure 4(a)). The increased stripping duration may be due to larger plated lithium during charging.

Cells' initial and final capacity (measured at C/10 rate) were compared to quantify the capacity loss due to cycling at 1C and 4C rates. The cell cycled at 1C rate showed minimal change in capacity, but the cells cycled at 4C rates showed a significant capacity loss for both loaded and unloaded cells. Measured discharge capacity loss (in %) is plotted as the function of the applied pressure for cells cycled at 4C in Figure 4(b) and shows that the capacity loss increases with increased applied pressure. Cells cycled under no external pressure show a capacity loss of about 13%, which increases with increasing pressure, leading to a capacity loss of about 23% at maximum pressure.

The changes in CE, capacity loss, and stripping duration with cycles suggest that cells charged at 4C rates are subjected larger amount of lithium plating during the initial cycles, and the lithium plating decreases with continued cycling. Application of compressive loads leads to reduced CE, larger capacity losses, and longer duration of lithium stripping, suggesting that lithium plating is exacerbated with compressive loads. The increase in parasitic reactions in the pouch cell causes a reduction in cyclable lithium, thus decreasing the cell's lifetime [35].

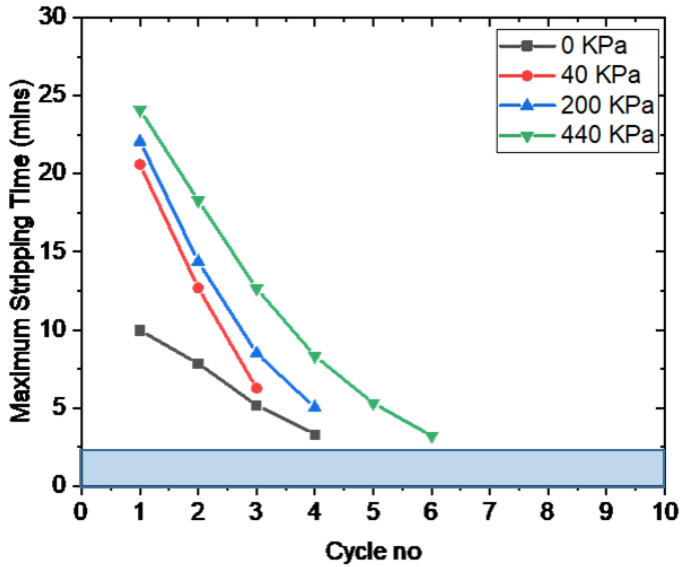

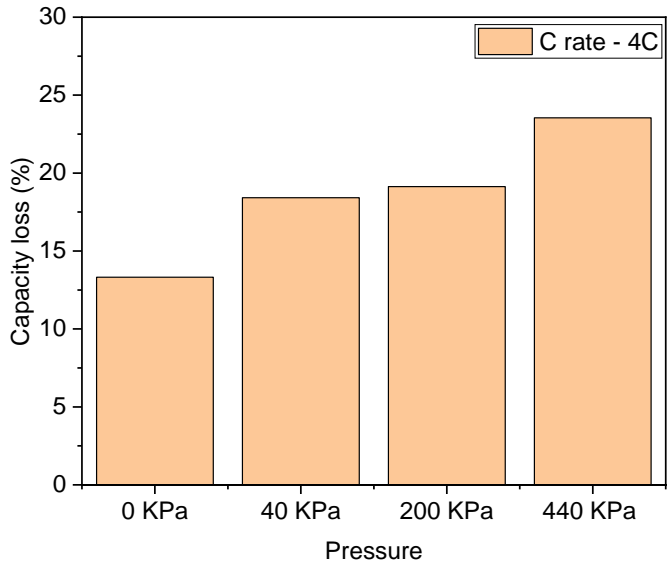

Figure 4: (a) Time duration from the beginning of voltage relaxation until dv-dt minima vs. number of cycles, (b) Capacity loss(%) vs. pressure applied for 4C charging rate.

After the cycling experiments, the cells were discharged and safely disassembled to observe the anode surface morphology for plated lithium using electron paramagnetic resonance (EPR) and scanning electron microscopy. EPR is a powerful, sensitive technique to detect conductive electrons and can be utilized to quantify the morphology of metallic lithium deposits [36-38]. The EPR spectra for the anodes from unloaded cells (0 KPa (Black)) are compared to anodes from compressively loaded cells (440 KPa (red)) after cycling at 1C and 4C rates in Figure S4 (shown in supplementary information). The EPR spectra plot the derivative of power absorbed by the sample as a function of the applied magnetic field [36, 39]. The peak intensity, linewidth from peak to peak, and peak shape indicate the plated lithium's amount and morphology on the graphite anode[40, 41].

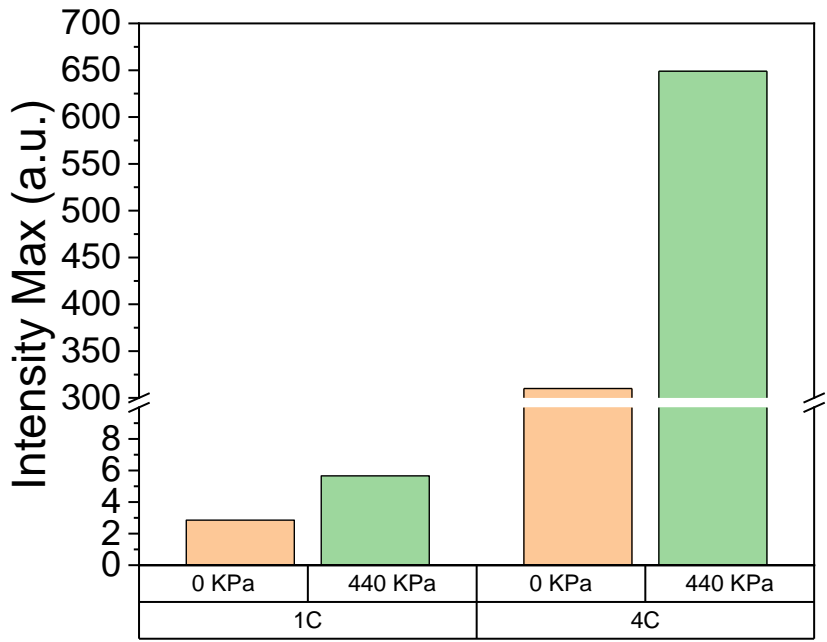

Figure 5: Comparison of Maximum Intensity obtained using EPR analysis at two charging rates, 1C and 4C, for two chosen pressure levels, 0 KPa and 440 KPa.

The anodes tested at a 1C charging rate (supplementary information figure S4(a)), the compressively loaded cells' peak intensity is approximately twice that of unloaded cells. The peak intensity in EPR spectra of graphite anodes indicates the presence of metallic or dead lithium isolated from the graphite substrate [38]. The peak-to-peak linewidth from anodes from loaded cells is greater than unloaded cells, indicating the presence of mossy lithium in the cells cycled under compressive loading. [42]

Figure S4(b) (supplementary information) shows the EPR spectra for the anodes from cells cycled at a 4C rate, and the peak intensity is almost two orders of magnitude greater than the peak intensity for cells charged with a 1C rate. The nominal peak-to-peak linewidth and Lorentzian lineshape of the spectra in Figure S4(b) indicate a dendritic morphology of the plated lithium in the anodes[42]. Additionally, for cells under 440 KPa high pressure, the amplitude shows a much stronger signal (almost doubled) than the cells at 0 KPa. The most prominent peak amplitude for cells charged at a 4C rate under 440 KPa pressure indicates the largest magnitude of metallic lithium dendritic deposits.

Figure 5 compares the maximum intensity peaks obtained from the EPR spectrum for chosen pressure levels for two charging rates. The results suggest that applying pressure increases the peak intensity, suggesting that higher lithium deposition may have occurred in cells under external mechanical pressure.

Backscattered electron-based SEM images of the anodes from cells cycled at 1C rate are shown in Figure 6. Low magnification images of the anodes for unloaded cells (Figure 6(a)) show graphite particles of various sizes with an average particle size of approximately 20μm. The voids between graphite particles show signs of localized agglomerations over the entire anode surface. In addition, the higher magnification micrograph, Figure 6(b), shows granular clumped agglomerations (highlighted in red circle) trapped between the graphite particles. The edges of the graphite particles in the BSE SEM micrographs appear brighter than the particle center, indicating a thicker SEI near the particle edges and a nonuniform distribution of the SEI layer over the anode surface.

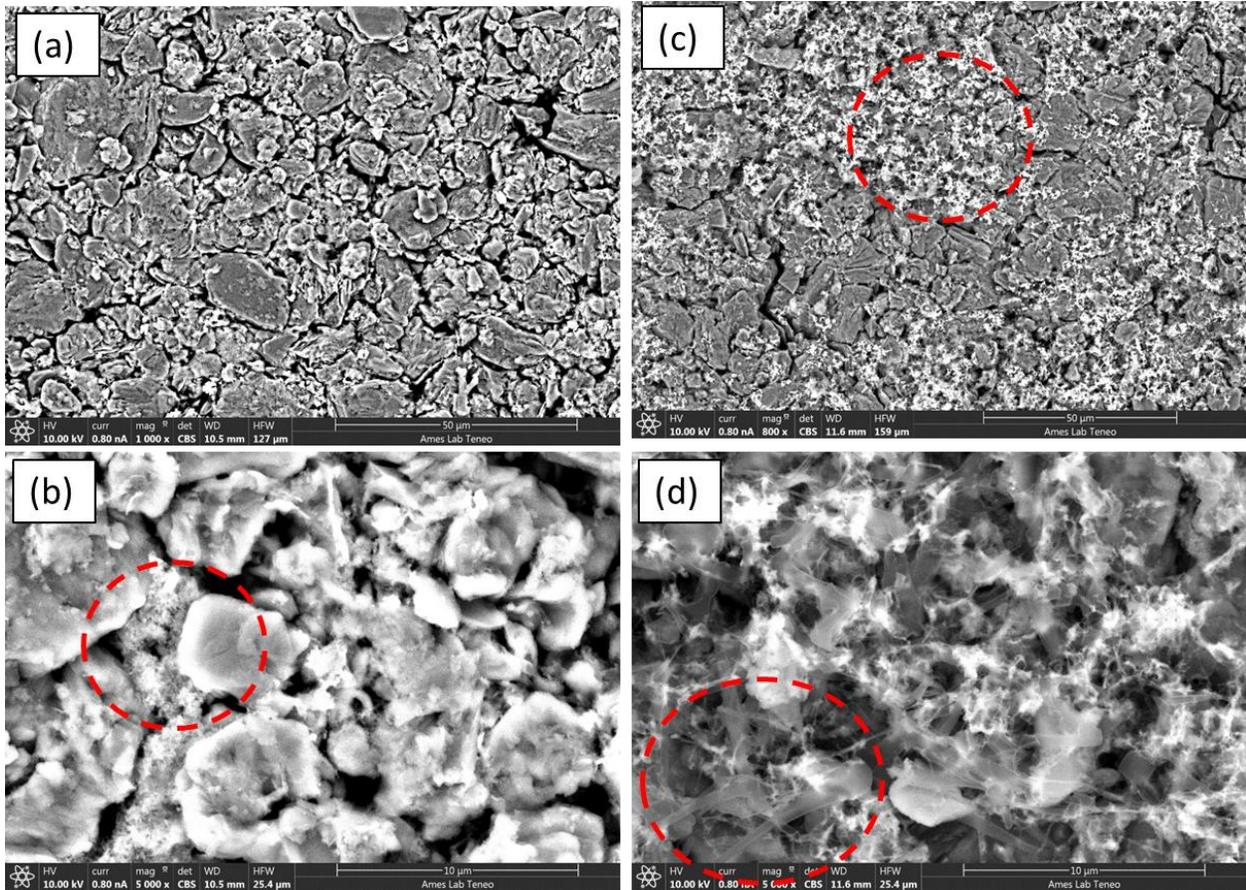

Figure 6: SEM BSE Images for graphite(anode) for 1C rate at different magnifications : (a)-(b) 0 KPa (no pressure) and (c)–(d) 440KPa.

Figure 6 shows the SEM BSE micrographs at two magnifications of the anode surface for cells cycled with 1C rate with two pressure conditions: (i) 0 KPa (shown in figures 6 (a) and 6(b)) and (ii) 440 KPa pressure (shown in figures 6(c) and 6(d)).

The anode of cells cycled under compressive pressure (shown in figures 6(c) and 6(d)) show similar graphite particle morphology as unloaded cells, but the interparticle voids are covered with filamentous deposits (highlighted in red circle). Comparison of the higher magnification SEM images (figures 6(b),6(d)) show that the filamentous deposits are more prominent on the cells cycled under compressive pressure. The shape/morphology of the rod-like structures is different from the clumped agglomerations in the unloaded cell (Figure 6(b)). The filamentous deposits are localized in the interparticle voids and do not cover the anode surface.

BSE SEM images of the anodes from cells cycled at a 4C rate are shown in Figure 7. Low magnification micrographs of the anodes cycled at 0 KPa (Figure 7(a)) exhibit a speckled pattern of deposits distributed over the whole surface. The micrograph of the anode from cells cycled under 440 KPa (Figure 7(c)) shows two distinct morphologies marked as region one and region two in Figure 7(c). In Region 1, the anode surface has a similar morphology as unloaded anodes and is covered with a speckled pattern of deposits. However, in region 2, the graphite particles are completely covered with a thick film of deposited material. Higher magnification SEM micrographs of the two distinct morphologies, region 1 (speckled deposits) and region 2 (completely covered deposits), are shown in Figures 7(b) and 7(d), respectively. Both images show that the anode surface is covered with filamentous deposits, but in Region 1, the deposits are localized in areas where the underlying graphite particles are visible. At the same time, the surface is completely covered with deposited material in region 2. The deposited material comprises agglomerates of intersecting filaments that are several microns long and have submicron diameters.

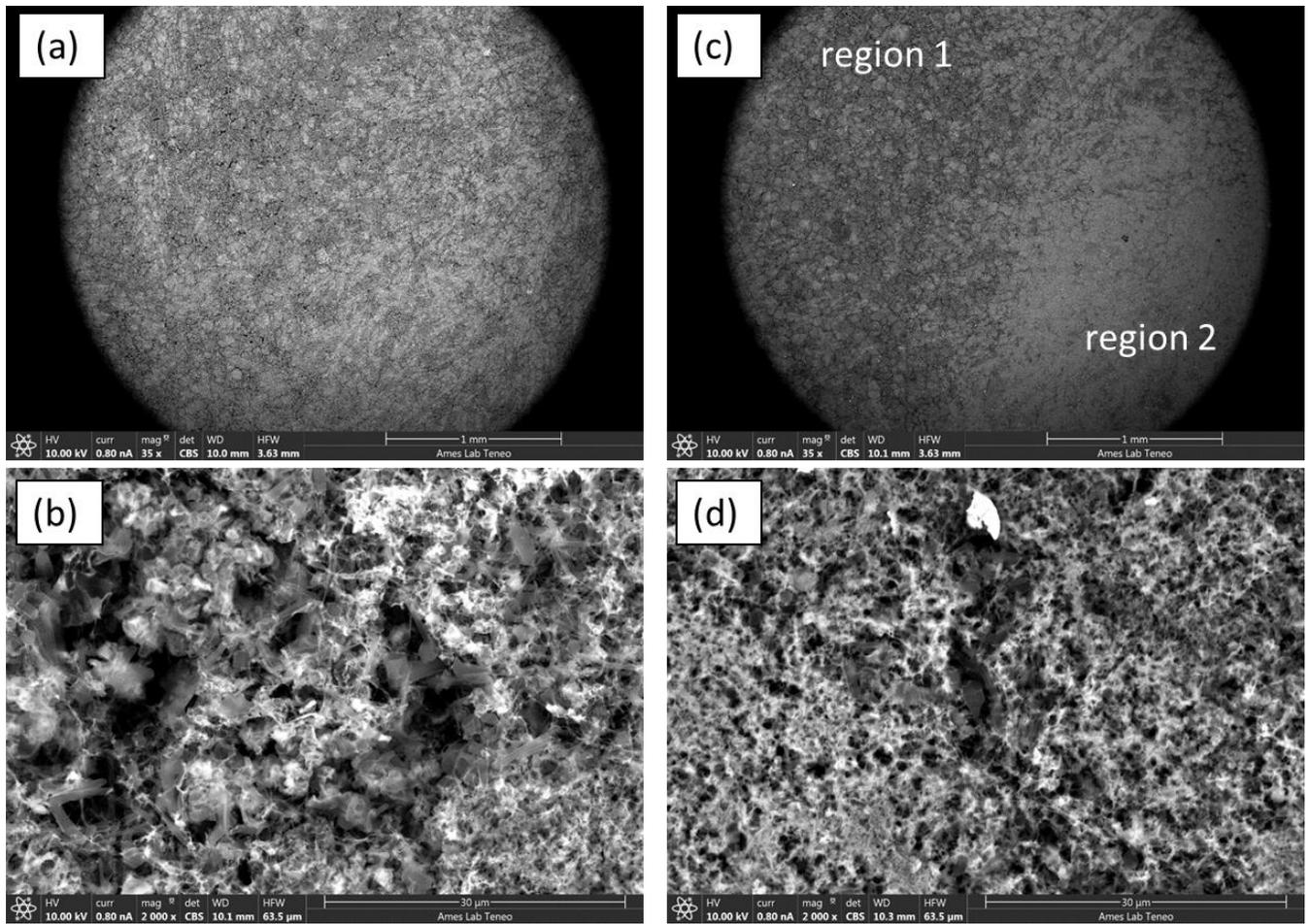

Figure 7: SEM micrographs for graphite electrode obtained from pouch cells charged under 4C rate for two pressure conditions (i) 0 KPa (pictures on the left) representing no pressure condition and 440KPa (pictures on the right) at two magnification levels (low magnification (a,c) and high magnification (b,d)) respectively.

Elemental composition maps obtained using Electron Dispersive X-ray Spectroscopy (EDS) of the anode surfaces from the cells tested at 1C 0KPa, 1C 440 KPa, 4C 0KPa, and dense film region (region 2) found on 4C 440 KPa are shown in Figure 8(a), (b), (c) and (d), respectively. The elemental map of the 1C 0 KPa anode surface shows that 99.3% of the surface is carbon-rich, with some amounts of phosphorus and fluorine in the interparticle voids. The anode surface of cells tested at 1C 440 KPa condition showed increased fluorine, phosphorus, and oxygen coverage of the surface and the associated drop in the exposed carbon, indicating an increased coverage of reaction products in the interparticle voids. The graphite particle surfaces are still exposed as carbon-rich surfaces. The elemental map of the 4C 0KPa anode surface shows that the surface is completely covered with reaction products with few and small carbon-rich spots where the underlying graphite particles are visible. The reaction product layer is rich in oxygen over the whole surface with speckle-shaped streaks rich in fluorine and phosphorus. An elemental map of anodes from cells tested at 4C 440 KPa shows that the surface is covered with similar reaction products. Detecting lighter elements like lithium is challenging using EDS, but the plated lithium on the anode reacts with electrolyte and is likely covered with a reaction layer with a similar composition [43]. Hence, the surface areas rich in oxygen, fluorine, and phosphorus may be associated with plated lithium material covered with reaction products. Cells with 4C 440 KPa show lower carbon percentage compared to 4C 0KPa. This decrease in carbon content in the surface characterization indicates that the surface is covered with a dense deposition layer.

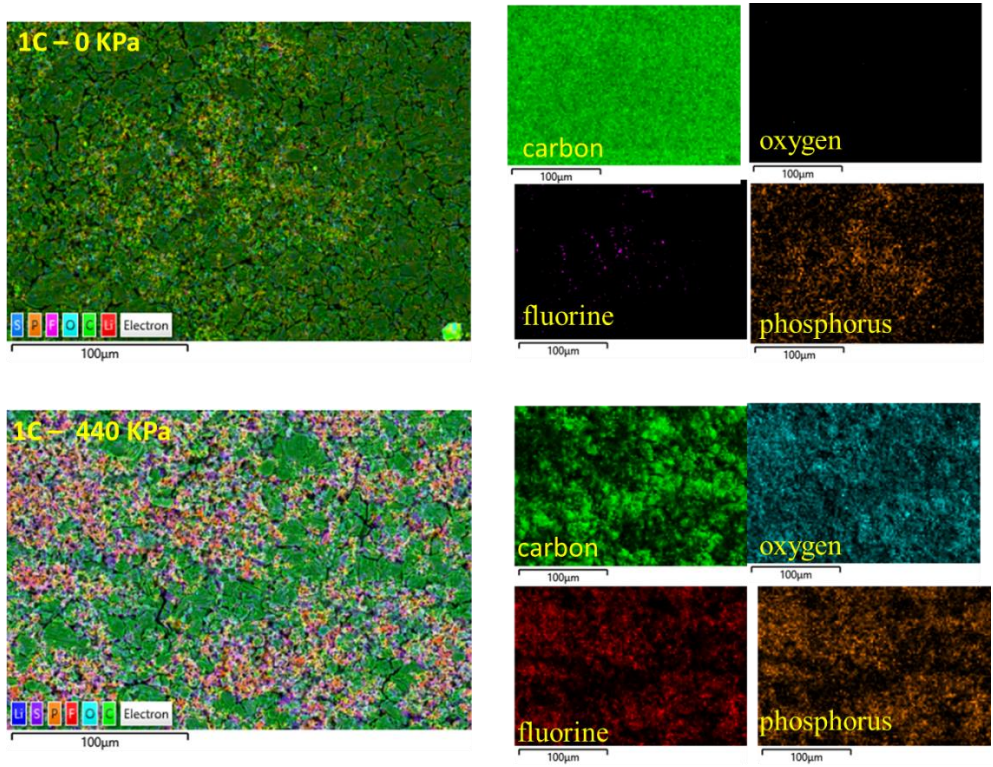

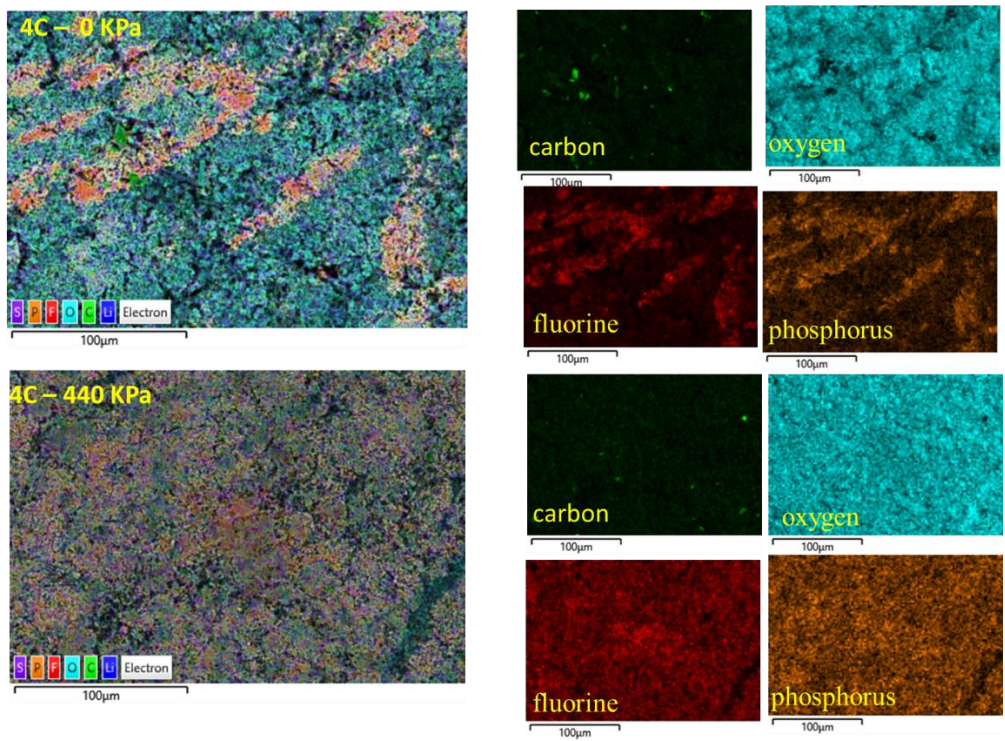

Figure 8: EDS maps for (a) 1C 0 KPa / no pressure applied, (b) 1C 440 KPa, (c) 4C 0 KPa / no pressure applied, and (d) 4C 440 KPa. The elemental maps show the atomic weight percentage (%) in yellow. (green color: carbon; blue color: oxygen; red color: fluorine; orange color; phosphorus)

## 4. DISCUSSION:

Cycling the pouch cell at low charging rates (1C) resulted in minimal capacity loss, no evidence of Li-stripping during relaxation after charging, and no influence of compressive loads on the charge-discharge response. The observations of no Li-plating-induced capacity loss are consistent with prior experiments on these cells where Li-plating losses were initiated at charging rates greater than 2C [44]. No change in charging and discharging characteristics was observed in the cells at 1C rate subjected to compressive load till 440 kPa. However, post-mortem examination of the anodes charge-cycled at 1C rate shows increased deposition of filamentous material in the pores between graphite particles and an increase in Li-associated EPR signal.

Cycling unloaded cells at charging rates (4C) above the plating threshold [44] resulted in large capacity losses and evidence of Li-stripping during relaxation after charging. Increasing the magnitude of mechanical compression resulted in increased capacity losses and an increase in Li-stripping duration. Observation of Li-stripping during relaxation and large capacity losses suggest that the loss of the lithium inventory is due to the parasitic reactions associated with Li-plating. Furthermore, applying compressive loads increases the amount of plated lithium during charging at 4C. During relaxation, a portion of the plated lithium is recovered through stripping and intercalation reactions. However, a significant portion gets electrically isolated or becomes dead lithium covered by an electrolyte reaction product layer. Post-mortem SEM and EDX examination of the anode surfaces shows regions of the filamentous deposits that are rich in electrolyte constituents, and the EPR measurements show a significant increase in lithium on the anode, indicating that the filamentous regions are dendritic dead lithium deposits covered with reaction product layers.

Arnold and co-workers [9, 11, 45-47] have demonstrated that compression-induced pore collapse defects in the polymeric separator lead to localized plating around the defects at low current densities. Increasing the current density results in uniform lithium plating around the defects. Peabody and Arnod [9] reported a loss in capacity on cells cycled at 0.5C rate subjected to compressive loads greater than 15 MPa and attributed the capacity loss to compressive pressure-induced pore closure in the polymeric separator. However, the capacity fades, and lithium plating observed in the current work cannot be attributed to the compressive pressure-induced defects.

The cells in the current experiments are subjected to significantly lower compressive pressure (440 kPa) than previous reports [9]. The low magnitude of compressive pressure results in primarily elastic deformation of the separator and may not result in pore collapse-associated defects in the LIB separator. Consequently, at low charging rates (1C), a negligible amount of lithium deposits are observed on the anodes, and there is a minimal loss of cell capacity during cycling, indicating a minimal loss in cyclable lithium inventory.

The capacity fades, lithium plating, and dead lithium deposits are only observed at charging rates above the plating threshold. The distribution pattern of the dead lithium deposits at fast charging conditions changes with the application of mechanical pressure. In the unloaded cells, there is a speckled pattern of deposits over the anode surface. The compressively loaded anodes have two morphologically distinct regions of dead lithium deposits – one similar to unloaded cells with speckled deposits and another region completely covered with filamentous deposits.

The non-uniformity of the lithium plating may have developed progressively over the ten charge-discharge cycles due to the combined influence of anode surface roughness, plated lithium asperities, and compressive loading. The columbic efficiency for the first cycle at 4C in both unloaded and loaded cells was of similar magnitude (as shown in Fig 2(b)), suggesting that both unloaded and loaded cells may have a similar pattern of lithium plating after the first cycle. The plated lithium asperities plated on the high portions of the rough anode surface could make contact with the separator layer. The compressive loading of this rough interface would result in localization of the contact and high stresses on the asperities near the surface peaks and may lead to the formation of pore closure defects in these regions. Subsequent cycling at the high charging rates would suppress lithium plating underneath the pore closure defects but enhance plating in the areas not under contact loading. Therefore, the combination of the lithium plating asperities induced surface roughness and compressive loading may have resulted in enhanced lithium plating and nonuniform distribution of plated lithium on cycled cells.

## 5. CONCLUSIONS:

We report the influence of nominally uniform compressive pressure on fast charging-induced capacity loss in LCO/G pouch cells. Applying compressive pressure till 440 KPa did not influence cell performance when cells cycled below the plating threshold. Cycling the cells above the plating threshold resulted in large capacity losses and evidence of Li-stripping during relaxation after charging. Increasing the magnitude of mechanical compression resulted in increased capacity losses and an increase in Li-stripping duration. EPR analysis of the anodes showed that anodes from the cells cycled above the plating threshold had large lithium dendritic deposits, and the magnitude of dendritic deposits increased with mechanical compression. Morphological and chemical characterization of the plated anodes indicated surfaces covered with speckled patterns of dead dendritic lithium deposits on the unloaded cells, while the compressively loaded cells had nonuniform patterns of lithium deposits – some areas with similar speckled patterns as unloaded cells while the other areas that were completely covered with dendritic lithium deposits. The results suggest that even low magnitudes of compressive pressure applied during fast charging of cells may result in significantly high lithium plating-associated capacity losses and pose enhanced risks to safe performance.

## ACKNOWLEDGEMENTS

This work was supported by the Department of Energy, Laboratory Directed Research and Development program at Ames Laboratory. Ames Laboratory is operated for the U.S. Department of Energy by Iowa State University of Science and Technology under Contract No. DE-AC02-07CH11358.

Supplementary Information:

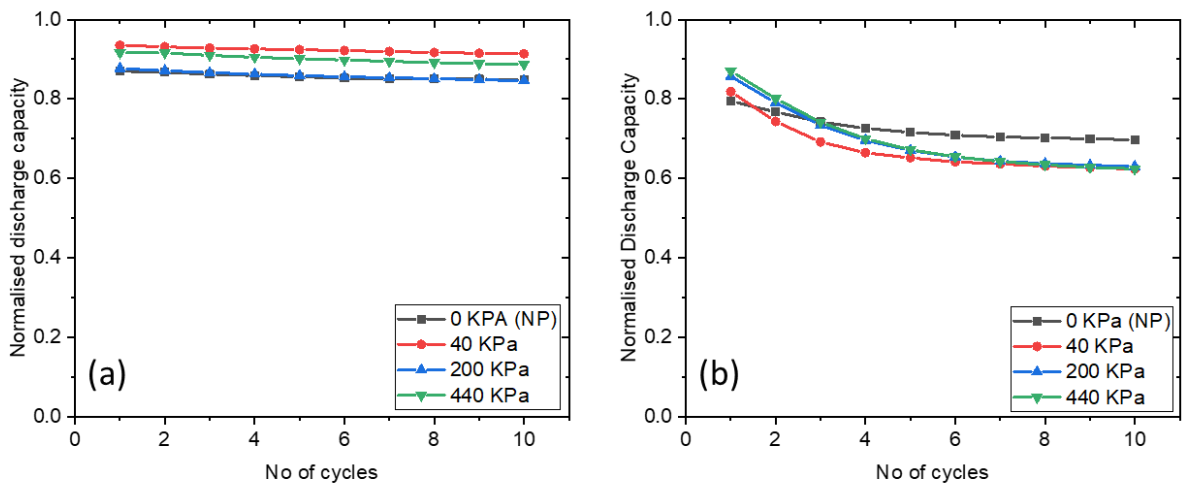

Figure S1: Normalized Discharge capacity for charging rate (a) 1C and (b) 4C

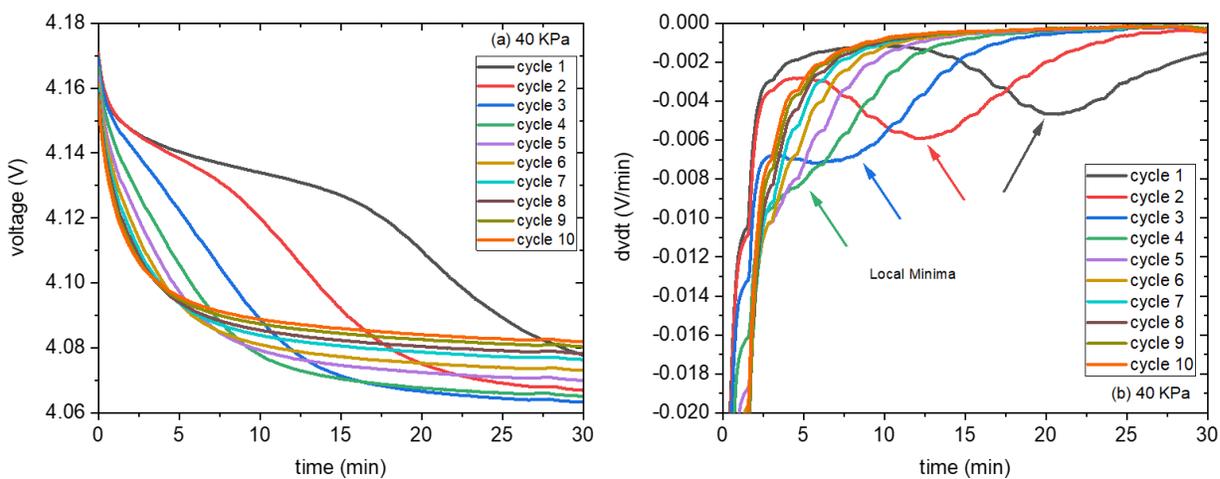

Figure S2: (a) Voltage during relaxation and (b) differential voltage vs. time for (a) 40 KPa

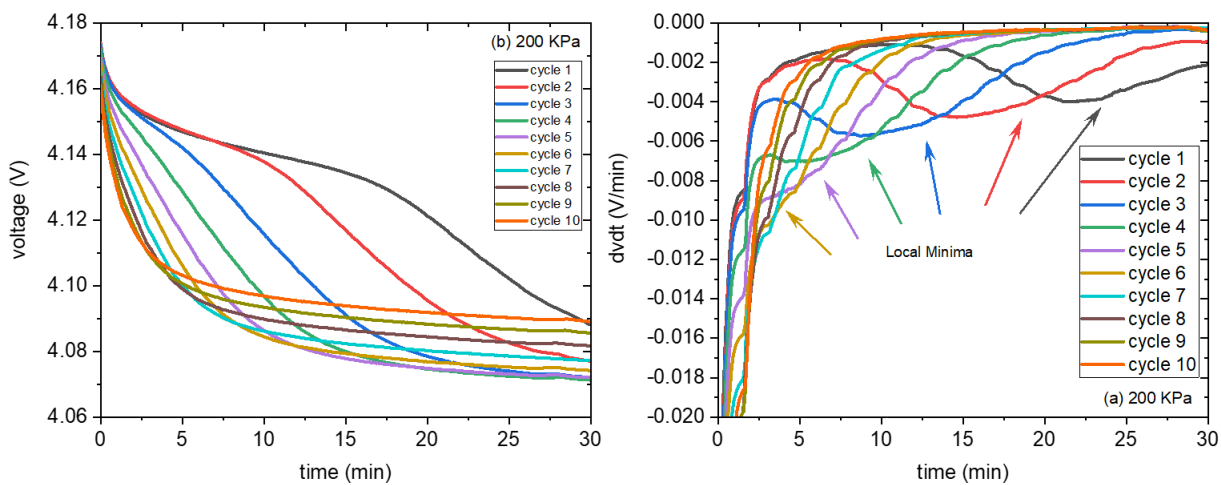

Figure S3: (a) Voltage during relaxation and (b) differential voltage vs. time for (a) 200 KPa

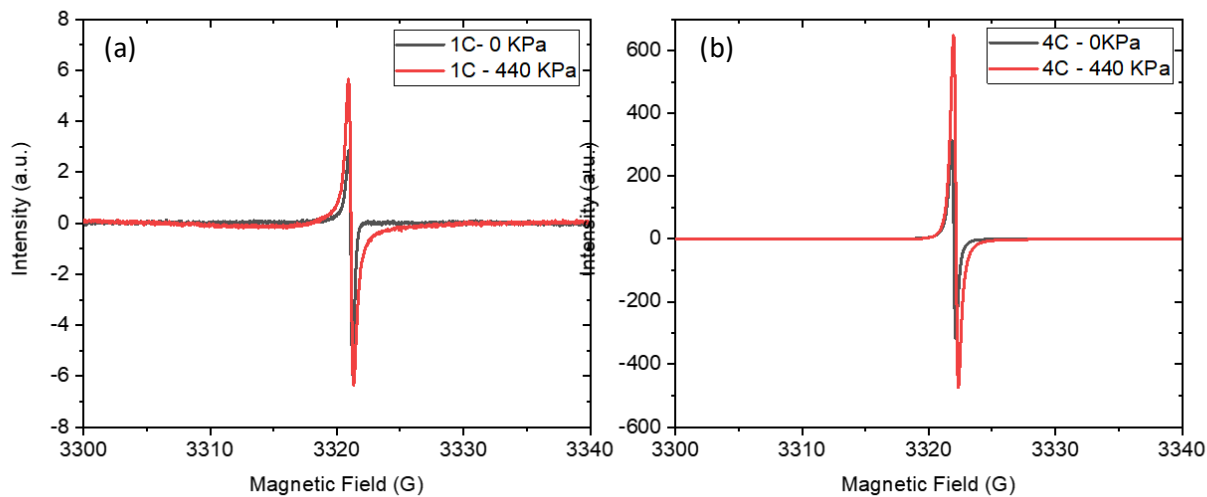

Figure S4: First derivative EPR signal obtained for cells cycled at two different C rates, (a) 1C and (b) 4C, at two pressure levels. The black spectra show the baseline representing cells at 0KPa pressure, and the spectra in red color show the cells at 440 KPa, presenting the effect of mechanical pressure for the chosen C rate.